# Statistical Estimates of Coordinate Error Circles for LAT-detected GRBs

Carl W. Akerlof and Fang Yuan

*Randall Laboratory of Physics, University of Michigan, Ann Arbor, Michigan 48109-1040*

## INTRODUCTION

The GLAST mission will explore a unique aspect of gamma-ray bursts, the component of radiation with photon energies that can exceed 10 GeV. The history of GRB research over the past ten years has demonstrated the importance of multiwavelength observations. With space missions such as HETE-II and Swift, this has become significantly easier with the advent of precise localizations for most events made possible by detectors operating at ~ 10 KeV energies. The GLAST burst monitor (GBM) has an angular resolution of the order of 10°, a field impractically large for prompt optical follow-up. Bursts detected by the Large Area Telescope (LAT) will be considerably better localized, providing an opportunity for prompt optical burst observations that will lead to detections with arc-second errors. To better plan a program of observations with our ROTSE-III telescopes, we have attempted to estimate the LAT localization errors for ensembles of detected gamma-ray photons. These estimates, independent of the more detailed Monte Carlo estimates of the GLAST team, may be useful for others interested in observing the afterglows of GLAST-detected bursts. We are indebted to Neil Gehrels and David Band for information about specific details of the GLAST mission.

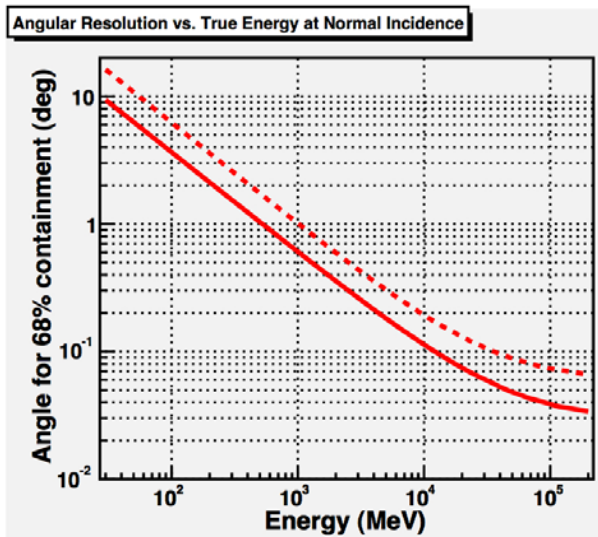

**FIGURE 1.** Angular resolution as a function of photon energy. The solid line shows the PSF for the thin converter layers; the dashed line corresponds to the thick layers.

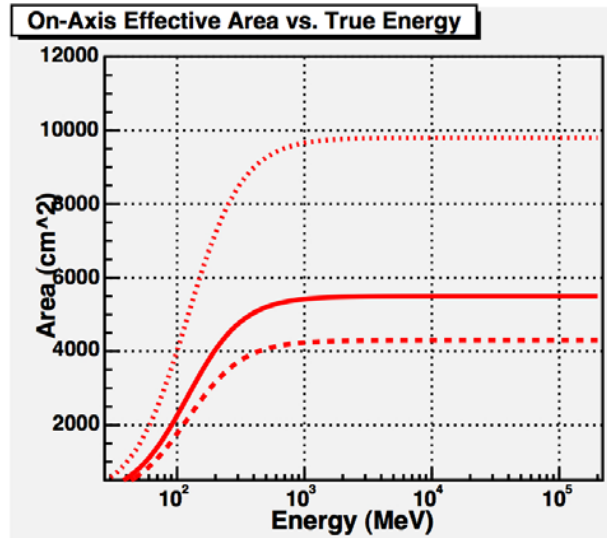

**FIGURE 2.** Effective detection area as a function of photon energy. The solid line shows the effective area for the thin converter layers; the dashed line corresponds to the thick layers; the dotted line is the sum of the two.

## LAT CHARACTERISTICS

The LAT tracking detector, TKR, spans a geometric area of 2.19 m$^2$ divided into 16 independent tower modules. Each tower contains alternating layers of tungsten foils and x-y planes of silicon strip ionization detectors. At the upstream end of the tower, 12 layers of 3% $X_0$ W converter foils and Si strips comprise the thin section followed by 4 layers of 18% $X_0$ foils and strips for the thick section. Two more x-y layers of Si strips complete the electron tracking. Two distributions characterize the LAT detector response to incident photons, the point spread function, PSF, and the effective detector area. The graphs shown above were obtained from the GLAST Web site:

http://www-glast.slac.stanford.edu/software/IS/glast_lat_performance.htm

## GRB DETECTION RATES AND SPECTRAL CHARACTERISTICS

GLAST is exploring a new frontier of high energy gamma-ray astronomy so there are considerable uncertainties in the rates of events, fluences, and spectral energy distributions. An estimate of the distribution of high energy spectral indices can be gleaned from the BATSE data[1]. Figure 3 below shows the distribution of power-law indices, β, for the brightest periods of a number of GRBs with photon energies ranging up to ~ 2 MeV. The distribution peaks at β ≈ -2.3 with most of the sample points lying between -3 and -2. As will be seen shortly, the LAT GRB localization error is essentially determined by the highest energy photon that is detected, strongly favoring bursts with β = -2 or greater. The extragalactic background light (EBL) will begin to kill gamma-ray photons above a few tens of GeV so observed spectra cannot follow such power laws indefinitely. If β ≥ -2 really characterizes the source spectrum, one is faced with the unphysical prediction of an unbounded fluence. Thus, any predictions about the GRB spectral energy distribution should be treated with some caution.

The GLAST team[2] has predicted the annual rate of LAT-detected GRBs as shown in Figure 4. The graph below is based on BATSE measurements and incorporates the cuts believed necessary for good event reconstruction. For the reasons given above, these estimates are subject to considerable error. A more conservative estimate[3] is one event per month with greater than 100 photons and one event per year with greater than 1000 photons. Since the probability of promptly observing a GRB at any given ground location is about 7%, finding optical counterparts of these events will require patience and fortitude.

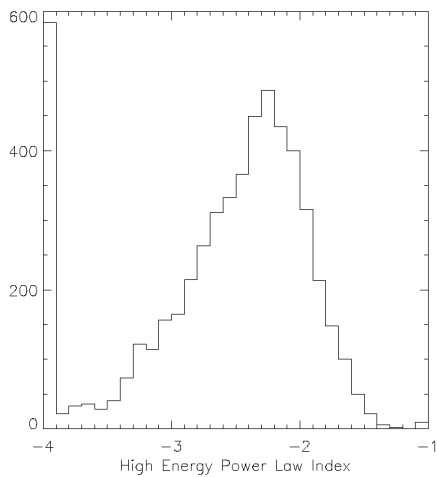 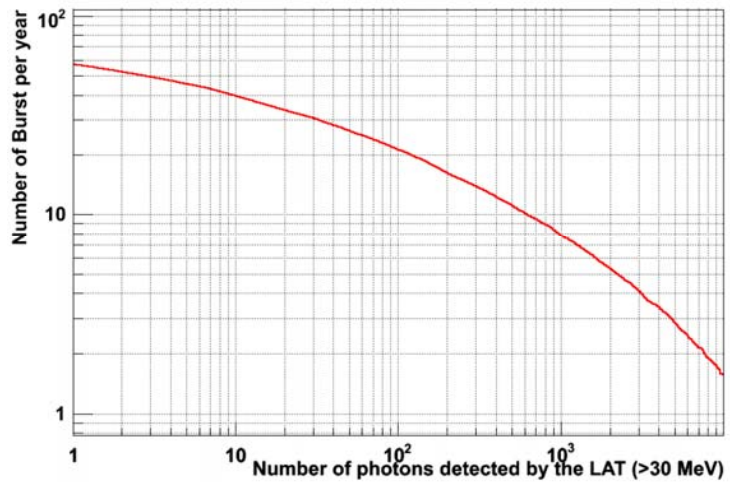

**FIGURE 3.** The differential distribution for the high energy power-law index, β. This plot was taken from Preece et al[1].

**FIGURE 4.** Number of GRBs per year detected by the LAT. 21 events per year are expected to exceed 100 photons. This data was provided by Neil Gehrels[2].

# CALCULATION OF GRB LOCALIZATION ERRORS

For an ensemble of detected photons, the angular error of the weighted celestial coordinates is given by:

$$\sigma_{GRB} = \left[ \sum_{i=1}^{N_\gamma} (\sigma_{68}(E_i))^{-2} \right]^{-\frac{1}{2}}$$

Thus, the only critical task is to compute the appropriate sum over the detected energy spectrum. The intrinsic GRB spectrum is assumed to be a power-law with a spectral index in the range, $-3 \leq \beta \leq -2$ while the detection efficiency is given by Figure 2. The GLAST team was not able to divulge the analytic expressions for the curves shown in Figures 1 and 2 so some reverse engineering was required to find the functional forms. Photoshop® was used to manually digitize the location of points along the curves and axes while the Solver tool in Excel® was used to find suitable fits. The accuracy of the fits is better than the line widths on the plots. For the PSF function, the results are:

$$\sigma_{68}(E) = \left(\frac{E_0}{E}\right)^\delta \sigma_0 + \sigma_1$$

where $E_0$ = 100 MeV, $\delta$ = 0.7956, $\sigma_0$ = {3.5384°, 6.1043°} and $\sigma_1$ = {0.0241°, 0.0482°} for {thin, thick} sections.

The effective detector area turned out to have the simple form:

$$A_{eff}(E) = \frac{C}{1 + (E/E_c)^{-2}}$$

with $E_c$ = 119.51 MeV and C = {5506.6 cm$^2$, 4298.4 cm$^2$} for {thin, thick} sections.

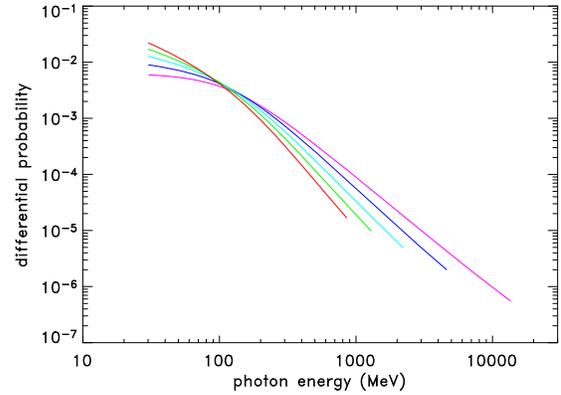

**FIGURE 5.** The differential spectrum of LAT-detected photons for various values of the spectral index in the range $-3 \leq \beta \leq -2$. The colors are matched to $\beta$ as follows: {-2.00, -2.25, -2.50, -2.75, -3.00}.

This was convenient because integrations over finite ranges can be expressed by analytic functions for spectral power-law indices that can be represented by n/4 with n an integer. We chose the set, $\beta$ = {-2.00, -2.25, -2.50, -2.75, -3.00}. This permits a straight-forward generation of Monte Carlo ensembles of photon energies for fixed photon count by finding the inverse of the cumulative probability distribution function. The resulting spectra are shown in Figure 5.

# ESTIMATED GRB LOCALIZATION ERRORS

The GRB localization errors computed by the method just described are shown below in Figures 6 and 7. For comparison, vertical lines show the bounds for some typical instruments that have been used for rapid observations of GRB afterglows. A number of effects have been ignored such as finite energy resolution of the calorimeter, degradation of the PSF for off-axis photons and the influence of PSF non-Gaussian tails. All of these will make the localization error larger. For this reason, dotted lines show the effects of a reasonably optimistic degradation factor of 1.5.

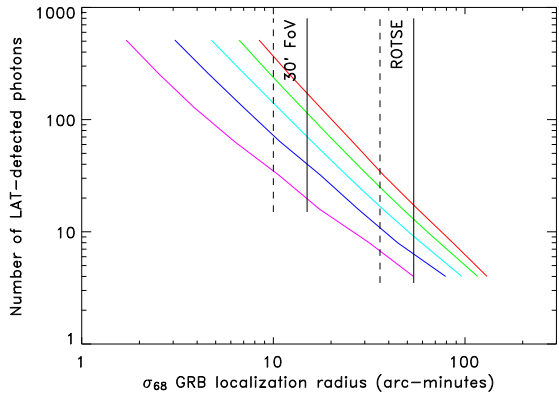 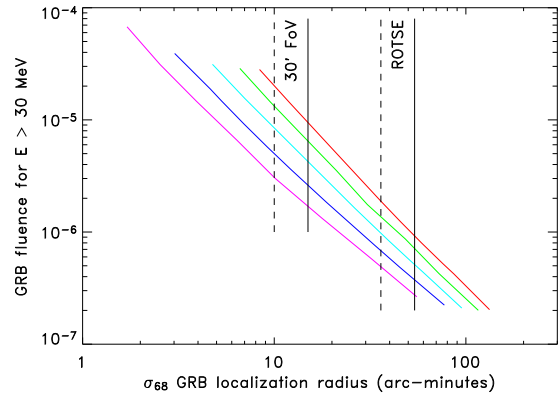

**FIGURE 6.** The number of required LAT-detected photons as a function of the median GRB localization error. The five lines indicate the behavior for β = {-2.00, -2.25, -2.50, -2.75, -3.00}. The vertical lines show the cutoffs for instruments with 30´ diameter FoV and ROTSE-III.

**FIGURE 7.** The required GRB fluence for E > 30 MeV as a function of the median GRB localization error. The five lines indicate the behavior for β = {-2.00, -2.25, -2.50, -2.75, -3.00}. The vertical lines show the cutoffs for instruments with 30´ diameter FoV and ROTSE-III.

## CONCLUSIONS

Extensive multiwavelength observations of LAT-detected GRBs will only be possible if ground-based instruments can identify these events efficiently. The dozen or so events per year with 100 photons or more could be found by instruments with fields-of-view with 15´ diameters for β = -2.00 and 24´ diameters for β = -2.25. As shown in Figure 8 below, this requirement eliminates 61% (for β = -2.00) to 76% (for β = -2.25) of the observatories that have been reporting GCN observations recently. Without the wider FoV systems such as RAPTOR, Super-LOTIS and ROTSE, the total discovery rate of GRB afterglows may be restricted to 3-4 events per year. If Nature is uncooperative, this could be considerably less. With a FoV of 110´ × 110´ and global coverage, ROTSE is much better matched to a broader range of spectral indices and fluences.

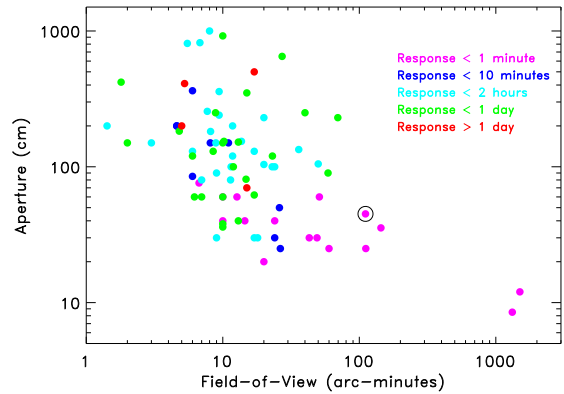

**FIGURE 8.** Scatter plot of 82 optical facilities producing GCN messages[4] during the period August 1, 2005 through August 1, 2006. The color code designates the minimum response time for the particular system. The parameters for the ROTSE-III instruments are identified by the circumscribed circle. The aperture and field-of-view were obtained from online documentation or relevant publications.

## ACKNOWLEDGEMENTS

This work has been supported by NASA grants NNG-04WC41G and NNG-06G190G. We gratefully acknowledge the many conversations and E-mails exchanged with the GLAST team.

## REFERENCES

1. Preece, et al., ApJ Suppl. 126, 19-36 (2000).
2. Neil Gehrels, private communication (2006).
3. David Band, private communication (2007).
4. Quimby, McMahon and Murphy, astro-ph/0312314 (2003).